%% file: _main.tex
\input{_constants}
\arxiv % \review OR \arxiv OR \cameraready

\pdfoutput=1
\documentclass[10pt,twocolumn,letterpaper]{article}
\input{cvpr_header}

\unless\ifarxiv \myexternaldocument{_supplementary} \fi

\begin{document}
%% TITLE
\title{\paperTitle}
\author{\authorBlock}
%\maketitle
\twocolumn[{%
\renewcommand\twocolumn[1][]{#1}%
\maketitle
\includegraphics[width=1\linewidth]{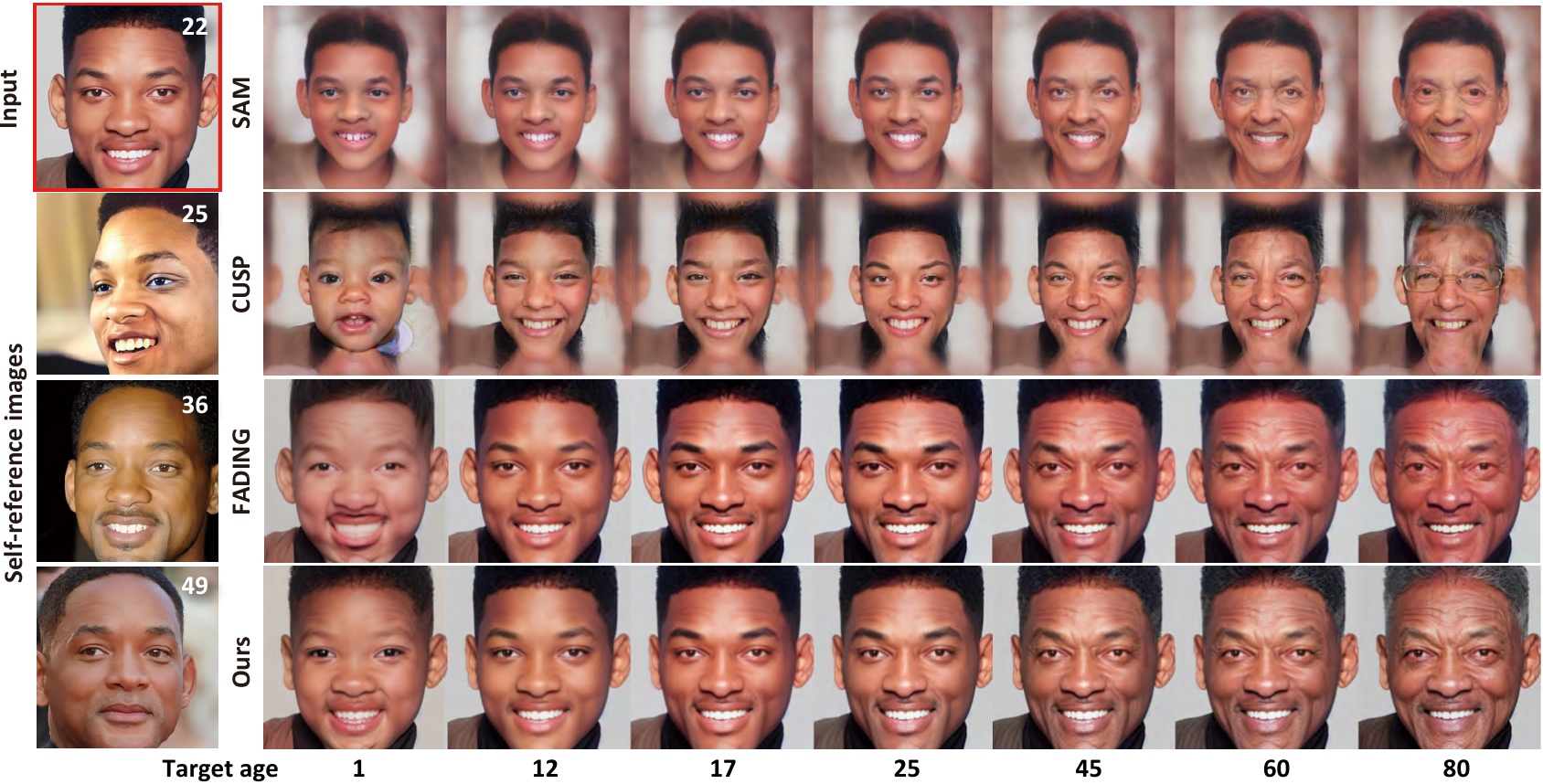}
%\vspace{-2em}
\captionof{figure}{Our method applies personalized age transformation to the input facial image (top left) using a few (3-5) self-reference images (left). The number on each image in the left column is the age estimated by an age estimator.\vspace{2em}}
\label{fig:teaser}
}]
\input{00_abstract}
\input{01_intro}
\input{02_related}
\input{03_method}
\input{04_experiment}
\input{10_conclusion}

{\small
\bibliographystyle{ieeenat_fullname}
\bibliography{11_references}
}

\ifarxiv \clearpage \appendix \input{12_appendix} \fi

\end{document}

%% file: _constants.tex
\def\paperID{0000}
\def\confName{CVPR}
\def\confYear{2025}

\def\paperTitle{SelfAge: Personalized Facial Age Transformation\\Using Self-reference Images}

\def\authorBlock{
    Taishi Ito%$\thanks{Equal contribution} 
    \qquad
     Yuki Endo %\footnotemark[1]
     \qquad
    Yoshihiro Kanamori \\
    University of Tsukuba \\
    {\tt\small itohlee14@gmail.com, \{endo, kanamori\}@cs.tsukuba.ac.jp}
}

\newif\ifreview 
\newif\ifarxiv \newcommand{\arxiv}{\arxivtrue}
\newif\ifcamera 
\newif\ifrebuttal 

%% file: cvpr_header.tex
\ifreview \usepackage[review]{cvpr} \fi
\ifarxiv \usepackage[pagenumbers]{cvpr} \fi
\ifrebuttal \usepackage[rebuttal]{cvpr} \fi
\ifcamera \usepackage{cvpr} \fi

\input{_macros}  % Add packages to _macros.tex

\usepackage{xr-hyper}

\makeatletter
\newcommand*{\addFileDependency}[1]{
  \typeout{(#1)}
  \@addtofilelist{#1}
  \IfFileExists{#1}{}{\typeout{No file #1.}}
}

\makeatother
\newcommand*{\myexternaldocument}[1]{
    \externaldocument{#1}
    \addFileDependency{#1.tex}
    \addFileDependency{#1.aux}
}

\definecolor{cvprblue}{rgb}{0.21,0.49,0.74}
\usepackage[pagebackref,breaklinks,colorlinks,citecolor=cvprblue]{hyperref}
\usepackage[capitalize]{cleveref}
\crefname{section}{Sec.}{Secs.}
\crefname{table}{Table}{Tables}
\crefname{figure}{Fig.}{Figs.}

\ifarxiv \crefname{appendix}{App.}{Apps.}
\else \crefname{appendix}{Suppl.}{Suppls.} \fi

%% file: _macros.tex
\usepackage{graphicx}	
\usepackage{amsmath}	
\usepackage{amssymb}	
\usepackage{booktabs}
\usepackage{times}
\usepackage{microtype}
\usepackage{epsfig}
\usepackage[table,xcdraw,dvipsnames]{xcolor}
\usepackage{caption}
\usepackage{float}
\usepackage{placeins}
\usepackage{color, colortbl}
\usepackage{stfloats}
\usepackage{enumitem}
\usepackage{tabularx}
\usepackage{xstring}
\usepackage{multirow}
\usepackage{xspace}
\usepackage{url}
\usepackage{subcaption}
\usepackage{xcolor}
\usepackage[hang,flushmargin]{footmisc}

\ifcamera \usepackage[accsupp]{axessibility} \fi

\usepackage{comment}
\usepackage{booktabs}
\usepackage{multirow}
\usepackage{graphicx}
\usepackage{xfrac}
\usepackage{placeins}
\usepackage[T1]{fontenc}
\usepackage{lmodern}
\usepackage{type1cm}
\usepackage{diagbox}
\usepackage{pifont}% http://ctan.org/pkg/pifont

% ...

\ifarxiv  \fi

\newcommand{\R}[1]{{%
    \textbf{%
        \ifstrequal{#1}{1}{\textcolor{red}{R#1}}{%
        \ifstrequal{#1}{2}{\textcolor{blue}{R#1}}{%
        \ifstrequal{#1}{3}{\textcolor{magenta}{R#1}}{%
        \ifstrequal{#1}{4}{\textcolor{teal}{R#1}}{%
                           \textcolor{cyan}{R#1}%
        }}}}%
    }%
}}

%% file: 00_abstract.tex
\begin{abstract}
% Abstract goes here.
Age transformation of facial images is a technique that edits age-related person's appearances while preserving the identity.
Existing deep learning-based methods can reproduce natural age transformations; 
however, they only reproduce averaged transitions and fail to account for individual-specific appearances influenced by their life histories.
In this paper, we propose the first diffusion model-based method for personalized age transformation.
Our diffusion model takes a facial image and a target age as input and generates an age-edited face image as output.
To reflect individual-specific features, we incorporate additional supervision using \textit{self-reference images}, which are facial images of the same person at different ages.
Specifically, we fine-tune a pretrained diffusion model for personalized adaptation using approximately 3 to 5 self-reference images.
Additionally, we design an effective prompt to enhance the performance of age editing and identity preservation.
Experiments demonstrate that our method achieves superior performance both quantitatively and qualitatively compared to existing methods. The code and the pretrained model are available at \href{https://github.com/shiiiijp/SelfAge}{\color{magenta}{https://github.com/shiiiijp/SelfAge}.}
\end{abstract}

%% file: 01_intro.tex
% To insert a figure: \input{figs/template}
% Or table: \input{tables/template}

\section{Introduction}
Age transformation of facial images is a task that aims to reproduce age-related appearance changes while preserving an individual's identity.
With advances in deep learning, realistic age editing has become possible using generative adversarial networks (GANs)~\cite{or2020lifespan,alaluf2021only,gomez2022custom} and diffusion models~\cite{li2023pluralistic,chen2023face}.
Existing methods learn typical age transformation from datasets containing facial images of diverse individuals. However, such typical age transformation often harms identity preservation and does not necessarily reflect the person-specific appearance transition resulting from the person's life history. 

To overcome this limitation, we propose the first diffusion model-based method for personalized age transformation that reflects individual characteristics (see Figure~\ref{fig:teaser}). To this end, we incorporate a few additional facial images of the target individual at different ages, \textit{self-reference images}, as supplementary supervision.
Specifically, inspired by Identity-Preserving Aging (IDP)~\cite{banerjee2023identity}, we fine-tune a pretrained latent diffusion model (LDM)~\cite{rombach2022high} to learn identity information from self-reference images while capturing age transformations from an age-labeled facial dataset~\cite{jiang2021talk}. 
However, unlike IDP, which generates facial images for a specified rough age group, our framework aims to edit existing facial images to integer target ages. 

During inference, our fine-tuned model generates a face image at a specified target age given an input facial image. 
Following the approach of Face Aging via Diffusion-based Editing (FADING)~\cite{chen2023face}, we first embed the input image into the latent space of the pretrained model using Null-text Inversion~\cite{mokady2023null}.
Next, we apply Prompt-to-Prompt~\cite{hertz2022prompt}, a method that edits images using paired prompts corresponding to pre- and post-edit states, to generate the age-edited output from the embedded representation.
We refine the prompt design to further enhance the performance of face aging and identity preservation. 

In summary, our main contributions are as follows:
\begin{enumerate}
    \item The first diffusion model-based method for personalized age transformation,
    \item Refinement of the regularization set to specify integer ages instead of age groups,
    \item Special adaptor to preserve a person's identity while avoiding overfitting on training data, and
    \item Careful prompt design for more accurate age transformation.
\end{enumerate}
We conduct extensive experiments to demonstrate that, compared to existing age editing methods, our method improves age editing accuracy while maintaining identity preservation. 

%% file: 02_related.tex
\section{Related Work}
\label{sec:related}

\paragraph{Age transformation.}
Many studies have explored controlling various facial attributes, including age, by manipulating latent variables in GANs~\cite{shen2020interpreting,wu2021stylespace,patashnik2021styleclip,parihar2022everything,huang2023adaptive,nitzan2022large,harkonen2020ganspace}.
For example, Shen et al.~\cite{shen2020interpreting} demonstrated that facial age could be edited by shifting latent variables along the normal directions of hyperplanes that separate attributes.
Huang et al.~\cite{huang2023adaptive} improved upon this by moving latent variables in multiple directions for a single attribute, enabling more natural age editing.
However, these methods allow for increasing or decreasing age but do not support specifying a precise target age.
Conversely, target age editing has been achieved within GAN-based image-to-image translation frameworks~\cite{or2020lifespan,alaluf2021only,gomez2022custom,yao2021high,DBLP:journals/vc/ItoEK23}.
Nevertheless, these GAN-based methods often struggle to preserve identity during age transformations.

Recently, diffusion models have gained significant attention, leading to the development of attribute editing methods~\cite{baumann2024continuous,kwon2022diffusion,li2023pluralistic,chen2023face}.
FADING~\cite{chen2023face} is an age editing method based on a pretrained LDM~\cite{rombach2022high}.
Specifically, it fine-tunes LDM on an age-labeled dataset to specialize the model for age editing.
During inference, the input image is embedded into the model's latent space using Null-text Inversion~\cite{mokady2023null}, and Prompt-to-Prompt~\cite{hertz2022prompt} is applied to modify only age-related regions.
However, these existing methods do not sufficiently consider individual variations in age progression and regression.

\paragraph{Personalized image synthesis.}

The task of adapting an image generative model to a specific concept is known as personalization.
Personalized image synthesis has been explored with
both GAN-based methods~\cite{roich2022pivotal,nitzan2022mystyle,qi2024my3dgen,zeng2023mystyle++} and diffusion-based methods~\cite{gal2022textual,ruiz2022dreambooth,kumari2023multi}.
Many of these approaches fine-tune pretrained models so that generated images become close to a small set of reference images.
IDP~\cite{banerjee2023identity} is a personalized age transformation method fine-tuned using self-reference images and diverse facial images~\cite{jiang2021talk}. 
However, because IDP is for generating new images but not for editing them, the composition and facial expression of the generated face image differ from those of the target person. Moreover, IDP restricts age input to predefined coarse categories (i.e., age groups). In contrast, our method can edit an existing image of the target person with integer target ages.

Concurrently, Qi et al.~\cite{DBLP:journals/corr/abs-2411-14521} proposed a personalized facial aging method. Their method is GAN-based, whereas our method is diffusion-based. Their work is a preprint at the time of our submission, and the source code is not publicly available. Direct comparison is left as future work. 

%% file: 03_method.tex
\section{Method}
\label{chap:method}

\begin{figure*}[t]
  \centering
  \includegraphics[width=\linewidth]{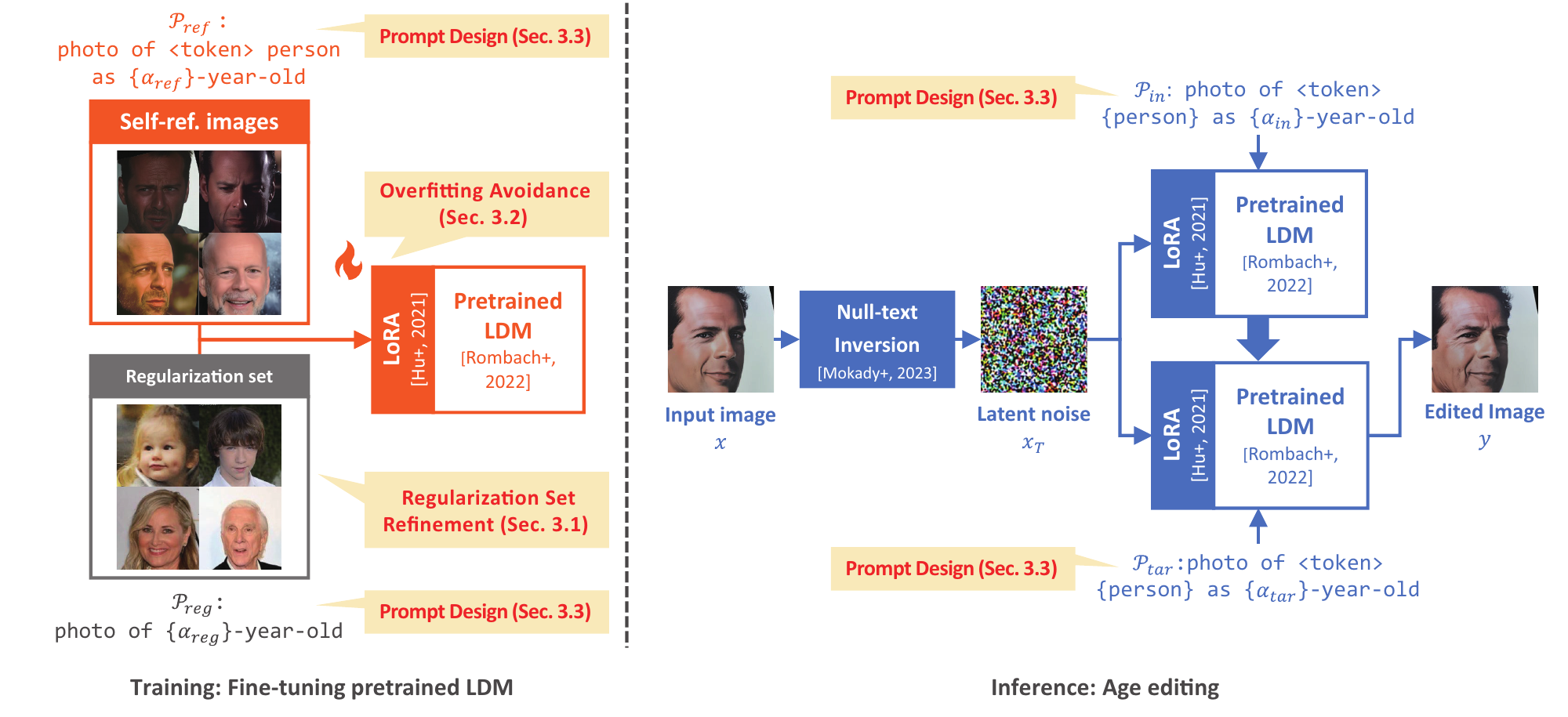}
  \caption{
Overview of our method. In the training phase, we fine-tune a pretrained diffusion model~\cite{rombach2022high} using a refined regularization set (see Section~\ref{sec:int_age}) and self-reference images labeled with integer ages. We employ LoRA~\cite{hu2021lora} to avoid overfitting on these images (see Section~\ref{sec:lora}). In the inference phase, from input image $x$, we first obtain a latent representation $x_T$ using Null-text Inversion~\cite{mokady2023null} and apply Prompt-to-prompt~\cite{hertz2022prompt} with original age $\alpha_\mathit{in}$ and target age $\alpha_\mathit{tar}$ to generate age-edited image $y$. We carefully design the text prompts $\mathcal{P}_\mathit{ref}$, $\mathcal{P}_\mathit{reg}$, $\mathcal{P}_\mathit{in}$, and $\mathcal{P}_\mathit{tar}$ for more accurate age transformation (see Section~\ref{sec:prompt_design} and Table~\ref{tab:prompt}).
  }
  \label{fig:our_method}
\end{figure*}

Our method aims to edit a given facial image $x$ to a target age $\alpha_\mathit{tar} \in \mathbb{Z}$. 
To capture personalized characteristics and enable identity-preserving age editing, our method incorporates additional information in the form of a few (3-5) self-reference images $\mathcal{X}=\{x_1,x_2,\ldots,x_M\}$ (where $M$ is the number of images) of the same individual, which differ from $x$.

Figure~\ref{fig:our_method} illustrates an overview of our method.
Our method is based on IDP~\cite{banerjee2023identity} and fine-tunes a pretrained LDM using self-reference images and prompts $\mathcal{P}_\mathit{ref}$.
This process embeds the target individual into the generative manifold by associating it with a unique token \texttt{$\langle$token$\rangle$}.
Simultaneously, it learns human age progression and regression from a regularization set, which consists of age-labeled facial images and prompts $\mathcal{P}_\mathit{reg}$.
For fine-tuning, we employ the same loss functions as IDP~\cite{banerjee2023identity}, which are the normalized temperature-scaled cross-entropy (NT-Xent) loss and the reconstruction losses for self-reference and regularization images. 
During inference, the input image is embedded into the diffusion model's latent space using Null-text Inversion~\cite{mokady2023null}.
Subsequently, age editing is performed using Prompt-to-Prompt~\cite{hertz2022prompt}, where a pair of prompts $\mathcal{P}_\mathit{in}$ and $\mathcal{P}_\mathit{tar}$ is provided to guide the transformation.
While our method is inspired by IDP and FADING, we found that a na\"ive combination of them does not show sufficient performance, and thus we introduce further improvements into our framework. 

\begin{table*}[t]
\small
  \centering
  \caption{
Comparison of input prompts used in IDP~\cite{banerjee2023identity}, FADING~\cite{chen2023face}, and our method. 
\texttt{$\langle$token$\rangle$} represents a token associated with an individual, and \texttt{$\langle$age group$\rangle$} denotes an age category.
$\alpha_\mathit{in}$, $\alpha_\mathit{tar}$, $\alpha_\mathit{ref}$, and $\alpha_\mathit{reg}$ are integer ages of the input image, target, self-reference image, and regularization image, respectively.
Additionally, \texttt{\{person\}} is dynamically replaced based on gender, input image age, and target age to ensure proper semantic alignment.
  }
  \label{tab:prompt}
  \begin{tabular}{llll}
    \hline
    \multicolumn{1}{c}{Prompt}
    & \multicolumn{1}{c}{IDP~\cite{banerjee2023identity}}
    & \multicolumn{1}{c}{FADING~\cite{chen2023face}}
    & \multicolumn{1}{c}{Ours} \\
    \hline
    $\mathcal{P}_\mathit{ref}$ & 
    \texttt{\begin{tabular}{l}
      photo of a $\langle$token$\rangle$ person
    \end{tabular}}
    &
    \texttt{\begin{tabular}{l}
      photo of $\alpha_\mathit{ref}$ year old person
    \end{tabular}}
    &
    \texttt{\begin{tabular}{l}
      photo of $\langle$token$\rangle$ person \\ as $\alpha_\mathit{ref}$-year-old
    \end{tabular}}
    \\
    \midrule
    $\mathcal{P}_\mathit{reg}$
    & 
    \texttt{\begin{tabular}{l}
      photo of a $\langle$age group$\rangle$
    \end{tabular}}
    &
    \texttt{\begin{tabular}{l}
      (N/A)
    \end{tabular}}
    &
    \texttt{\begin{tabular}{l}
      photo of person \\ as $\alpha_\mathit{reg}$-year-old
    \end{tabular}}
    \\
    \midrule
    $\mathcal{P}_\mathit{in/tar}$
    & 
    \texttt{\begin{tabular}{l}
      photo of a $\langle$token$\rangle$ person \\ as $\langle$age group$\rangle$
    \end{tabular}}
    &
    \texttt{\begin{tabular}{l}
      photo of $\alpha_\mathit{in/tar}$ year old \{person\}
    \end{tabular}}
    &
    \texttt{\begin{tabular}{l}
      photo of $\langle$token$\rangle$ \{person\} \\ as $\alpha_\mathit{in/tar}$-year-old
    \end{tabular}}
    \\
    \hline
  \end{tabular}
\end{table*}

\subsection{Regularization Set Refinement
}
\label{sec:int_age}

IDP~\cite{banerjee2023identity} only allows age editing within predefined age groups, such as teenager or elderly, and does not support integer age specification.
This limitation arises from the regularization set used in training.
The existing regularization set consists of 612 facial images sampled from the CelebA-Dialog dataset~\cite{jiang2021talk}.
CelebA-Dialog provides age group labels, categorizing each image into six groups (\texttt{child}, \texttt{teenager}, \texttt{young adults}, \texttt{middle-aged}, \texttt{elderly}, \texttt{old}) with 102 images per group.
To overcome this limitation, our method reassigns integer age labels to the IDP regularization set using a pretrained age estimator, DEX~\cite{rothe2015dex}.
DEX is a neural network that predicts integer ages from 0 to 100 based on facial images.
By incorporating these integer labels, our method learns age transformations explicitly linked to numerical ages, enabling precise and flexible age editing for any target integer age.

Similarly to our method, FADING~\cite{chen2023face} fine-tunes a diffusion model using integer ages as supervision. Specifically, it assigns the median age of the corresponding age group as a label to 150 images sampled from the FFHQ-Aging dataset~\cite{or2020lifespan}.
In contrast, our method fine-tunes the model using a larger and more diverse set of images sampled evenly from existing age groups, with estimated integer ages as labels.
This enables our model to learn a broader and more detailed representation of age transformations.

\begin{figure}
  \centering
  \includegraphics[width=\linewidth]{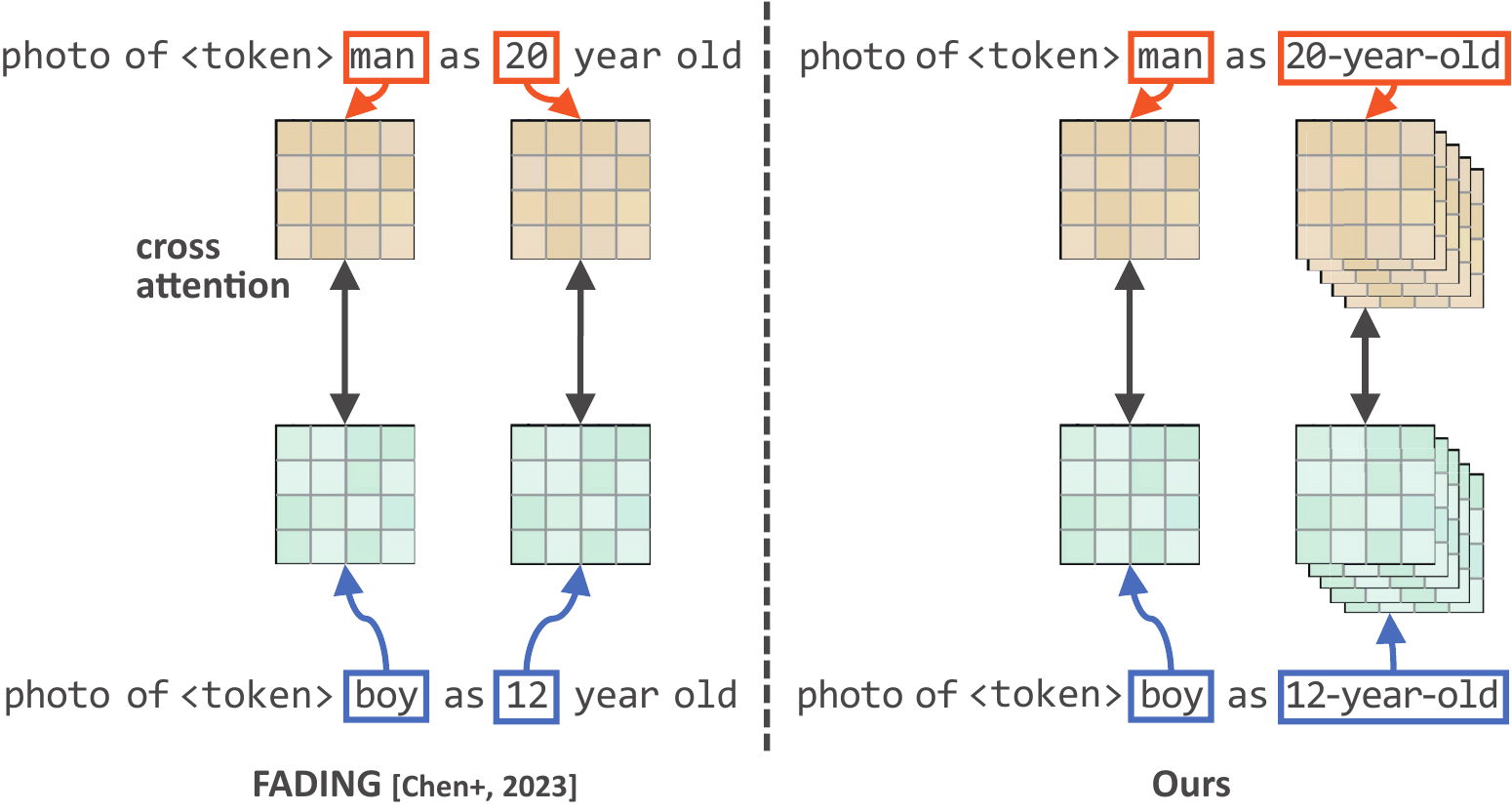}
  \caption{
  Difference in cross-attention value replacement between our method and FADING~\cite{chen2023face}. Our method represents age information as ``\texttt{$\alpha$-year-old}" and replaces cross-attention values corresponding to the person-describing noun (e.g., ``\texttt{man}'' or ``\texttt{boy}'') as well as the tokens ``\texttt{$\alpha$}", ``\texttt{-}", ``\texttt{year}", ``\texttt{-}", and ``\texttt{old}". In contrast, FADING represents age information as ``\texttt{$\alpha$ year old}" and replaces cross-attention values only for the person-describing noun and the token ``\texttt{$\alpha$}".
  }
  \label{fig:yearold}
\end{figure}

\subsection{Overfitting Avoidance with LoRA
}
\label{sec:lora}
Fine-tuning the entire U-Net, as done in IDP~\cite{banerjee2023identity} and FADING~\cite{chen2023face}, can lead to overfitting on self-reference images, resulting in unnatural age transformations.
To solve this problem, our method introduces Low-Rank Adaptation (LoRA)~\cite{hu2021lora}, which trains additional adapter layers instead of fine-tuning the entire model. 
Specifically, the pretrained weight matrix $W$ of U-Net is updated as $W' = W + \Delta W$, where $\Delta W\in \mathbb{R}^{d\times d}$ represents the learned adaptation.
Here, $\Delta W$ is approximated as a low-rank decomposition, $\Delta W=AB$, where $A\in \mathbb{R}^{d\times r}$ and $B\in \mathbb{R}^{r\times d}$, with natural integers $r \ll d$.

\subsection{Prompt Design
}\label{sec:prompt_design}

\begin{table}[t]
  \centering
  \caption{
  Word replacement in the prompts for our method. The placeholder \texttt{\{person\}} is replaced based on age $\alpha$ and gender, where it is substituted with ``\texttt{man}", ``\texttt{woman}", ``\texttt{boy}", ``\texttt{girl}", ``\texttt{baby}", or ``\texttt{elderly}", as appropriate.
  }
  \label{tab:person}
  \begin{tabular}{ccc}
    \toprule
    \multirow{2}{*}{Age} & \multicolumn{2}{c}{Gender} \\
    \cmidrule(lr){2-3}
    & Male & Female \\
    \midrule
    $\alpha<5$ & \multicolumn{2}{c}{\texttt{baby}} \\
    $5\le\alpha<15$ & \texttt{boy} & \texttt{girl} \\
    $15\le\alpha<65$ & \texttt{man} & \texttt{woman} \\
    $65\le\alpha$ & \multicolumn{2}{c}{\texttt{elderly}} \\
    \bottomrule
  \end{tabular}
\end{table}

\paragraph{Modification of age representation.
}
Our method utilizes a different prompt design from IDP~\cite{banerjee2023identity} and FADING~\cite{chen2023face} to effectively incorporate the personal identity token \texttt{$\langle$token$\rangle$} learned from fine-tuning with self-reference images. 
Table~\ref{tab:prompt} compares the input prompts used in each method.
Given these formulations, a na\"ive adaptation would combine elements from both approaches.
Specifically, \texttt{$\langle$token$\rangle$} is placed before the noun phrase, and the age expression follows, connected with ``\texttt{as}''. Thus, the prompts are structured as:
$\mathcal{P}_\mathit{ref}$: ``\texttt{photo of $\langle$token$\rangle$ person}'',
$\mathcal{P}_\mathit{reg}$: ``\texttt{photo of person as $\alpha_\mathit{ref}$ year old}'', and
$\mathcal{P}_\mathit{in/tar}$: ``\texttt{photo of $\langle$token$\rangle$ \{person\} as $\alpha_\mathit{in/tar}$ year old.}''

However, we found this na\"ive approach does not achieve sufficient performance because of the inappropriate age representation in a prompt (see Section~\ref{sec:eval_yearold}). 
In IDP, the model learns age information by associating facial images with a single word token \texttt{$\langle$age group$\rangle$} during regularization.
In contrast, FADING and our method specify integer ages using a three-word phrase, ``\texttt{$\alpha$ year old.}''
Additionally, in FADING, when applying Prompt-to-Prompt during inference, only the cross-attention values corresponding to ``\texttt{$\alpha$}'' are replaced.
However, because ``\texttt{year}'' and ``\texttt{old}'' also contribute to age representation, the cross-attention replacement may not correctly perform age editing. 
To address this issue, our method modifies the age expression to ``\texttt{$\alpha$-year-old}'', connecting the words with hyphens to ensure the model interprets the entire phrase as a single cohesive token.
Furthermore, we replace the cross-attention values corresponding to ``\texttt{$\alpha$}'', ``\texttt{-}'', ``\texttt{year}'', ``\texttt{-}'', and ``\texttt{old}'' to ensure proper age modification.
Figure~\ref{fig:yearold} illustrates the difference in cross-attention value replacement between our method and FADING. Our final prompts are shown in Table~\ref{tab:prompt}. 

\paragraph{Use of self-reference image age.}
As shown in $\mathcal{P}_\mathit{ref}$ in Table~\ref{tab:prompt}, IDP~\cite{banerjee2023identity} embeds identity information into the model using self-reference images but does not utilize any age-related information from those images.
In contrast, our method explicitly incorporates the age $\alpha_\mathit{ref}$ of self-reference images into the input prompt during training.
This encourages the model to disentangle identity features and age features.
Specifically, we modify the training prompt $\mathcal{P}_\mathit{ref}$ for self-reference images as 
``\texttt{photo of $\langle$token$\rangle$ person as $\alpha_\mathit{ref}$-year-old.}''
Here, \texttt{$\langle$token$\rangle$} corresponds to identity information, while $\alpha_\mathit{ref}$-year-old explicitly encodes age information.

\paragraph{Token replacement for extreme age.
}

To provide more precise guidance for inference, FADING~\cite{chen2023face} modifies the word \texttt{\{person\}} in the inference prompt based on the input image's gender and age, replacing it with ``\texttt{man}", ``\texttt{woman}", ``\texttt{boy}", or ``\texttt{girl}" as appropriate.
Our method further refines this approach to enhance performance, particularly for extreme age transformations, such as very young or very old ages. Specifically, as shown in Table~\ref{tab:person}, we modify the editing prompt $\mathcal{P}_\mathit{in/tar}$ by changing {person} based on the target age. When the target age is below 5, {person} is replaced with ``\texttt{baby}", and when the target age is 65 or older, it is replaced with ``\texttt{elderly}".

%% file: 04_experiment.tex
\section{Experiments}
\label{chap:experiments}

\begin{table*}[t]
  \centering
  \caption{
  Influence of the number of self-reference images. The \textbf{boldface} indicates the best scores while the \underline{underline} the second best scores.
  }
  \label{tab:refnum}
  \resizebox{\linewidth}{!}{
    \begin{tabular}{llccccccccccc}
      \toprule
      \multirow{2}{*}{Metric} & \multirow{2}{*}{Method} & \multicolumn{11}{c}{Target age} \\
      \cmidrule(lr){3-13}
      & & 1 & 5 & 8 & 12 & 17 & 25 & 35 & 45 & 60 & 80 & ALL \\
      \midrule
      \multirow{4}{*}{AGE~$\downarrow$} & Ours w/ 0 ref. & 8.92 & \underline{16.8} & 14.2 & \underline{9.80} & \textbf{11.0} & \textbf{11.4} & \underline{11.8} & 13.1 & \underline{7.28} & \textbf{6.52} & \textbf{11.1} \\
      & Ours w/ 1 ref. & \textbf{8.41} & \textbf{16.7} & \textbf{14.0} & 10.1 & 12.4 & 13.0 & 11.9 & \underline{12.2} & \textbf{6.95} & \underline{6.56} & \underline{11.2} \\
      & Ours w/ 3 ref. & 9.22 & 17.2 & \underline{14.1} & \textbf{9.54} & 12.3 & 12.4 & \textbf{11.7} & \textbf{11.9} & 7.42 & 7.07 & 11.3 \\
      & Ours w/ 5 ref. & \underline{8.89} & 17.8 & 14.4 & 10.3 & \underline{12.2} & \underline{12.3} & 12.2 & 12.5 & 7.48 & 7.31 & 11.5 \\
      \midrule
      \multirow{4}{*}{ID~$\downarrow$} & Ours w/ 0 ref. & \textbf{0.155} & \underline{0.124} & \underline{0.117} & \underline{0.106} & 0.0677 & 0.0656 & 0.0620 & 0.0680 & 0.0748 & 0.141 & 0.0981 \\
      & Ours w/ 1 ref. & 0.168 & 0.135 & 0.123 & 0.117 & 0.0698 & 0.0663 & 0.0623 & 0.0718 & 0.0757 & 0.134 & 0.102 \\
      & Ours w/ 3 ref. & \textbf{0.155} & \underline{0.124} & \textbf{0.115} & \textbf{0.105} & \textbf{0.0630} & \textbf{0.0545} & \textbf{0.0517} & \textbf{0.0613} & \textbf{0.0649} & \textbf{0.128} & \textbf{0.0923} \\
      & Ours w/ 5 ref. & \underline{0.156} & \textbf{0.122} & \textbf{0.115} & \underline{0.106} & \underline{0.0653} & \underline{0.0581} & \underline{0.0568} & \underline{0.0677} & \underline{0.0703} & \underline{0.129} & \underline{0.0946} \\
      \bottomrule
    \end{tabular}
  }
\end{table*}

\begin{table*}[t]
  \centering
  \caption{
    Quantitative comparison between our method and the existing methods. 
    }
  \label{tab:final}
  \resizebox{\linewidth}{!}{
  \begin{tabular}{llccccccccccc}
    \toprule
    \multirow{2}{*}{Metric} & \multirow{2}{*}{Method} & \multicolumn{11}{c}{Target age} \\
    \cmidrule(lr){3-13}
    & & 1 & 5 & 8 & 12 & 17 & 25 & 35 & 45 & 60 & 80 & ALL \\
    \midrule
    \multirow{4}{*}{AGE~$\downarrow$} & SAM~\cite{alaluf2021only} & 17.8 & \textbf{15.9} & \textbf{13.9} & \underline{11.8} & \textbf{8.74} & \textbf{7.79} & \textbf{10.8} & \underline{11.7} & 8.77 & \textbf{3.55} & \textbf{11.1} \\
    & CUSP~\cite{gomez2022custom} & 13.0 & 20.4 & 24.4 & 20.0 & 12.4 & \underline{10.6} & 12.2 & \textbf{10.2} & \textbf{6.37} & 7.11 & 13.7 \\
    & FADING~\cite{chen2023face} & \underline{11.7} & 21.7 & 20.3 & 15.4 & 16.3 & 15.5 & 15.9 & 14.7 & 8.86 & 7.72 & 14.8 \\
    & Ours & \textbf{9.22} & \underline{17.2} & \underline{14.1} & \textbf{9.54} & \underline{12.3} & 12.4 & \underline{11.7} & 11.9 & \underline{7.42} & \underline{7.07} & \underline{11.3} \\
    \midrule
    \multirow{4}{*}{ID~$\downarrow$} & SAM~\cite{alaluf2021only} & 0.270 & 0.268 & 0.266 & 0.264 & 0.261 & 0.261 & 0.264 & 0.266 & 0.271 & 0.271 & 0.266 \\
    & CUSP~\cite{gomez2022custom} & 0.208 & 0.204 & 0.208 & 0.201 & 0.195 & 0.145 & 0.140 & 0.146 & 0.167 & 0.283 & 0.190 \\
    & FADING~\cite{chen2023face} & \textbf{0.151} & \textbf{0.103} & \textbf{0.0987} & \textbf{0.0812} & \textbf{0.0603} & \underline{0.0606} & \underline{0.0638} & \underline{0.0714} & \underline{0.0885} & \textbf{0.121} & \textbf{0.0900} \\
    & Ours & \underline{0.155} & \underline{0.124} & \underline{0.115} & \underline{0.105} & \underline{0.0630} & \textbf{0.0545} & \textbf{0.0517} & \textbf{0.0613} & \textbf{0.0649} & \underline{0.128} & \underline{0.0923} \\
    \bottomrule
  \end{tabular}
  }
\end{table*}

\begin{figure*}[t]
  \centering
  \includegraphics[width=1.\linewidth]{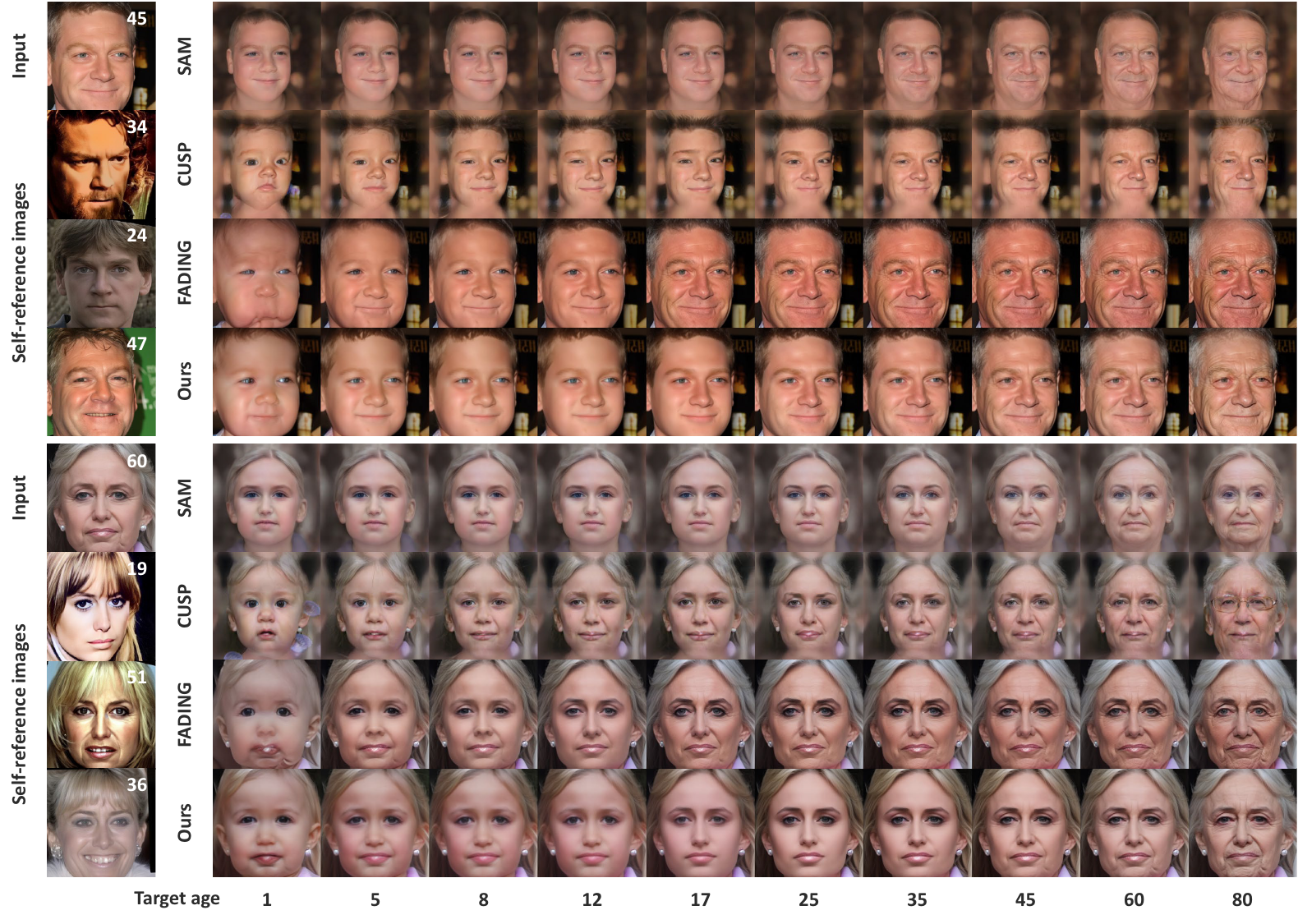}
  \caption{
  Qualitative comparison between our method and the existing methods~\cite{alaluf2021only,gomez2022custom,chen2023face}. The upper right numbers on the input and self-reference images show the ages estimated by the age estimator. 
  }
  \label{fig:qualitative}
\end{figure*}

\begin{table*}[t]
  \centering
  \caption{
  Quantitative comparison of our method with and without our refined regularization set. 
  }
  \label{tab:dexreg}
  \resizebox{\textwidth}{!}{
  \begin{tabular}{llccccccccccc}
    \toprule
    \multirow{2}{*}{Metric} & \multirow{2}{*}{Method} & \multicolumn{11}{c}{Target age} \\
    \cmidrule(lr){3-13}
    & & 1 & 5 & 8 & 12 & 17 & 25 & 35 & 45 & 60 & 80 & ALL \\
    \midrule
    \multirow{2}{*}{AGE~$\downarrow$} & Ours w/o refined reg. & 9.51 & 17.3 & 14.4 & 9.91 & \textbf{12.0} & \textbf{12.1} & \textbf{11.7} & \textbf{11.7} & 7.52 & 7.77 & 11.4 \\
    & Ours & \textbf{9.22} & \textbf{17.2} & \textbf{14.1} & \textbf{9.54} & 12.3 & 12.4 & \textbf{11.7} & 11.9 & \textbf{7.42} & \textbf{7.07} & \textbf{11.3} \\
    \midrule
    \multirow{2}{*}{ID~$\downarrow$} & Ours w/o refined reg. & 0.163 & 0.131 & 0.123 & 0.113 & 0.0731 & 0.0649 & 0.0626 & 0.0714 & 0.0752 & 0.129 & 0.101 \\
    & Ours & \textbf{0.155} & \textbf{0.124} & \textbf{0.115} & \textbf{0.105} & \textbf{0.0630} & \textbf{0.0545} & \textbf{0.0517} & \textbf{0.0613} & \textbf{0.0649} & \textbf{0.128} & \textbf{0.0923} \\
    \bottomrule
  \end{tabular}
  }
\end{table*}

\paragraph{Experimental settings. }
Our method was implemented using Python, PyTorch, and Diffusers.
For both training and inference, we used an NVIDIA RTX A6000 GPU.
The input image size was $224\times224$ pixels.
Our method utilized Stable Diffusion v1.5~\cite{rombach2022high} as the pretrained diffusion model. The batch size was set to 2, and training was performed for 800 iterations.
We used AdamW~\cite{loshchilov2017decoupled} as the optimizer with a learning rate of $1.0\times10^{-6}$.
The LoRA rank was $r=16$. 
Training took approximately 35 minutes, while inference required around 60 seconds for Null-text Inversion and 15 seconds for Prompt-to-Prompt.

\paragraph{Datasets.}
For a regularization set, we used 594 out of 612 images from CelebA-Dialog~\cite{jiang2021talk}, which were properly aligned for the age estimator DEX~\cite{rothe2015dex}.
For self-reference images, we constructed a dataset based on AgeDB~\cite{moschoglou2017agedb}.
AgeDB is a dataset consisting of 16,488 images of 568 celebrities collected from the Internet, with an average of 29 images per individual.
It is labeled by DEX with integer age labels ranging from 0 to 101.
However, AgeDB images have a relatively low resolution of $112\times112$ pixels and contain low-quality images, including grayscale ones.
To address this, we colorized the grayscale images using an existing method~\cite{kang2023ddcolor} and applied super-resolution~\cite{lin2024diffbir} to all images.
We used this dataset for both training and inference in our experiment. 
Specifically, we selected 20 individuals from the dataset (10 males and 10 females) and trained the model using a few self-reference images per individual.
For inference, we randomly selected five images of the corresponding individual for each model. 

\paragraph{Evaluation metrics.}
We used AGE and ID as evaluation metrics~\cite{alaluf2021only}.
AGE represents the accuracy of age editing and is calculated as the mean absolute difference between the estimated age of the output image, obtained from Face++~\cite{facepp}, and the target age. 
ID measures identity preservation and is computed as the average cosine similarity between the feature vectors of the input and output images using ArcFace~\cite{deng2019arcface}.

\subsection{
Influence of the Number of Self-reference Images
}
\label{sec:eval_refnum}

First, we investigated the influence of the number of self-reference images.
As shown in Table~\ref{tab:refnum}, our method demonstrates improvements in ID using a few self-reference images. 
Notably, using three self-reference images achieves the best ID score in total (ALL) while suppressing deterioration in AGE, indicating a more balanced and stable adaptation. We set the number of self-reference images $M=3$ for all subsequent experiments.

\subsection{Comparisons with Existing Methods}
\label{sec:comparison}

We compared our method with the GAN-based age editing methods, SAM~\cite{alaluf2021only} and CUSP~\cite{gomez2022custom}, as well as the diffusion-based method, FADING~\cite{chen2023face}.

\paragraph{Quantitative comparison.}
As shown in Table~\ref{tab:final}, our method demonstrates the second-best AGE score in total, following SAM.
Meanwhile, the GAN-based methods, SAM and CUSP, show significantly worse ID scores than our method.
While FADING performed the best in ID, it gets much worse in AGE than our method.
These results indicate that our method achieves accurate age editing without significantly degrading identity preservation.

\paragraph{Qualitative comparison.}
Figure~\ref{fig:qualitative} presents the qualitative results.
SAM exhibits significant identity changes due to its low inversion performance of input images and fails to perform adequate age regression for younger targets. 
CUSP, while effective at editing toward younger ages, tends to struggle with identity preservation.
Additionally, as seen in the second row of the lower example, aging transformations sometimes introduce unintended attributes, such as the addition of glasses due to attribute entanglement. 
FADING suffers from noticeable artifacts at younger ages and exhibits abrupt transitions around the 12-25 age range.
In contrast, our method successfully produces convincing age-editing results while preserving the distinct characteristics of the given self-reference images. 
More comparisons are shown in Appendix. 

\subsection{Ablation Studies}
This section validates the effectiveness of our improvements described in Sections~\ref{sec:int_age},~\ref{sec:prompt_design}, and~\ref{sec:lora}.
\label{sec:ablation}

\subsubsection{Regularization set refinement}
Table~\ref{tab:dexreg} presents a quantitative comparison of our method with and without regularization set refinement.
The results indicate improvements in both AGE and ID.
Notably, ID improves across all target ages, demonstrating the effectiveness of the regularization set.

\begin{table*}[t]
  \centering
  \caption{
  Quantitative comparison of our method with and without LoRA~\cite{hu2021lora}. 
  }
  \label{tab:lora_existence}
  \resizebox{\linewidth}{!}{
  \begin{tabular}{llccccccccccc}
    \toprule
    \multirow{2}{*}{Metric} & \multirow{2}{*}{Method} & \multicolumn{11}{c}{Target age} \\
    \cmidrule(lr){3-13}
    & & 1 & 5 & 8 & 12 & 17 & 25 & 35 & 45 & 60 & 80 & ALL \\
    \midrule
    \multirow{2}{*}{AGE~$\downarrow$} & Ours w/o LoRA & 9.68 & 17.5 & 16.4 & 10.6 & \textbf{9.45} & \textbf{10.4} & \textbf{9.12} & \textbf{10.7} & 9.36 & \textbf{5.74} & \textbf{10.9} \\
    & Ours & \textbf{9.22} & \textbf{17.2} & \textbf{14.1} & \textbf{9.54} & 12.3 & 12.4 & 11.7 & 11.9 & \textbf{7.42} & 7.07 & 11.3 \\
    \midrule
    \multirow{2}{*}{ID~$\downarrow$} & Ours w/o LoRA & 0.297 & 0.327 & 0.325 & 0.310 & 0.253 & 0.243 & 0.252 & 0.262 & 0.266 & 0.300 & 0.283 \\
    & Ours & \textbf{0.155} & \textbf{0.124} & \textbf{0.115} & \textbf{0.105} & \textbf{0.0630} & \textbf{0.0545} & \textbf{0.0517} & \textbf{0.0613} & \textbf{0.0649} & \textbf{0.128} & \textbf{0.0923} \\
    \bottomrule
  \end{tabular}
  }
\end{table*}

\begin{figure*}[t]
  \centering
  \includegraphics[width=\linewidth]{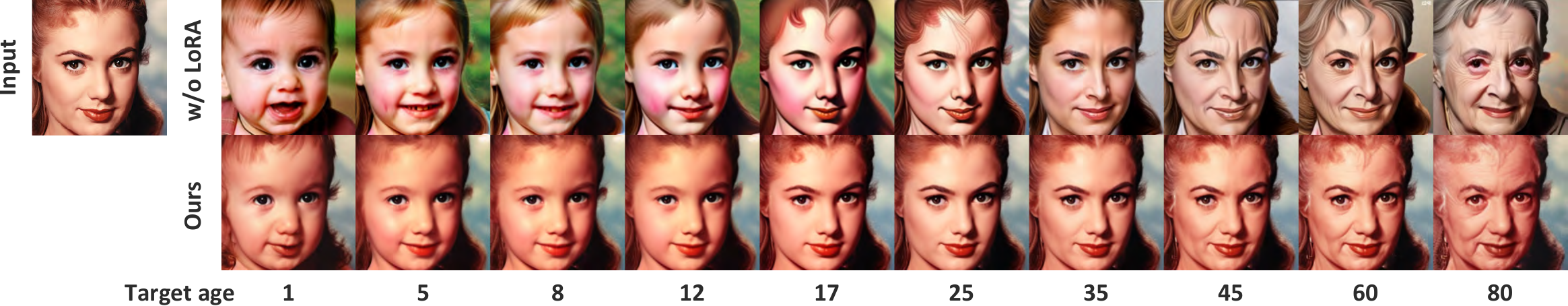}
  \caption{
    Qualitative comparison of our method with and without LoRA~\cite{hu2021lora}. 
  }
  \label{fig:lora_existence}
\end{figure*}

\subsubsection{
Overfitting avoidance with LoRA
}
We evaluated the impact of introducing LoRA into our method.
As shown in Table~\ref{tab:lora_existence}, incorporating LoRA improves ID across all target ages.
While our method without LoRA shows better AGE scores in some age ranges, it significantly sacrifices ID.
As shown in the qualitative results in Figure~\ref{fig:lora_existence}, without LoRA, the facial appearance is changed into a painterly style, leading to noticeable quality degradation of the output images.

\subsubsection{Prompt design}

\paragraph{
Modification of age representation.}
\label{sec:eval_yearold}

\begin{table*}[t]
  \centering
  \caption{
    Quantitative comparison of our method with and without hyphenation of age representation and the corresponding cross-attention replacement. 
    }
  \label{tab:yearold}
  \resizebox{\linewidth}{!}{
    \begin{tabular}{llccccccccccc}
      \toprule
      \multirow{2}{*}{Metric} & \multirow{2}{*}{Method} & \multicolumn{11}{c}{Target age} \\
      \cmidrule(lr){3-13}
      & & 1 & 5 & 8 & 12 & 17 & 25 & 35 & 45 & 60 & 80 & ALL \\
      \midrule
      \multirow{2}{*}{AGE~$\downarrow$} & Ours w/o hyphens & \textbf{7.65} & \textbf{16.6} & \textbf{13.9} & 9.68 & \textbf{8.57} & 12.9 & 15.1 & 12.7 & 7.71 & 7.70 & \textbf{11.2} \\
      & Ours & 9.22 & 17.2 & 14.1 & \textbf{9.54} & 12.3 & \textbf{12.4} & \textbf{11.7} & \textbf{11.9} & \textbf{7.42} & \textbf{7.07} & 11.3 \\
      \midrule
      \multirow{2}{*}{ID~$\downarrow$} & Ours w/o hyphens & 0.176 & 0.139 & 0.134 & 0.120 & 0.0797 & 0.0611 & 0.0604 & 0.0722 & 0.0791 & 0.135 & 0.106 \\
      & Ours & \textbf{0.155} & \textbf{0.124} & \textbf{0.115} & \textbf{0.105} & \textbf{0.0630} & \textbf{0.0545} & \textbf{0.0517} & \textbf{0.0613} & \textbf{0.0649} & \textbf{0.128} & \textbf{0.0923} \\
      \bottomrule
    \end{tabular}
  }
\end{table*}

We evaluated the impact of modifying the age representation prompt from ``\texttt{$\alpha$ year old}" to ``\texttt{$\alpha$-year-old}" and the corresponding cross-attention replacement. 
Table~\ref{tab:yearold} presents the quantitative comparison.
The results show that this modification allows our method to improve ID while maintaining AGE. Notably, ID preservation improved across all target ages.

\paragraph{
Use of self-reference image age.}
\begin{table*}[t]
  \centering
  \caption{
    Quantitative comparison of our method with and without using self-reference image age. 
    }
  \label{tab:refage}
  \resizebox{\linewidth}{!}{
    \begin{tabular}{llccccccccccc}
      \toprule
      \multirow{2}{*}{Metric} & \multirow{2}{*}{Method} & \multicolumn{11}{c}{Target age} \\
      \cmidrule(lr){3-13}
      & & 1 & 5 & 8 & 12 & 17 & 25 & 35 & 45 & 60 & 80 & ALL \\
      \midrule
      \multirow{2}{*}{AGE~$\downarrow$} & Ours w/o $\alpha_\mathit{ref}$ & \textbf{8.69} & \textbf{16.6} & \textbf{13.9} & \textbf{9.44} & \textbf{10.6} & \textbf{11.9} & 11.8 & \textbf{11.6} & 7.57 & \textbf{6.94} & \textbf{10.9} \\
      & Ours & 9.22 & 17.2 & 14.1 & 9.54 & 12.3 & 12.4 & \textbf{11.7} & 11.9 & \textbf{7.42} & 7.07 & 11.3 \\
      \midrule
      \multirow{2}{*}{ID~$\downarrow$} & Ours w/o $\alpha_\mathit{ref}$ & 0.163 & 0.129 & 0.119 & 0.112 & 0.0675 & 0.0639 & 0.0605 & 0.0699 & 0.0769 & 0.139 & 0.100 \\
      & Ours & \textbf{0.155} & \textbf{0.124} & \textbf{0.115} & \textbf{0.105} & \textbf{0.0630} & \textbf{0.0545} & \textbf{0.0517} & \textbf{0.0613} & \textbf{0.0649} & \textbf{0.128} & \textbf{0.0923} \\
      \bottomrule
    \end{tabular}
  }
\end{table*}

Table~\ref{tab:refage} presents the performance changes resulting from incorporating the age of self-reference images as input.
The results indicate that, although overall AGE slightly deteriorates, ID improves across all target ages.
This suggests that by incorporating the age of self-reference images, our method effectively disentangles age information from identity, enabling more efficient identity learning.

\paragraph{
Token replacement for extreme age. 
}
\label{sec:eval_baby}

\begin{table}[t]
  \centering
  \caption{
    Quantitative comparison with and without token replacement for extreme age. 
    }
  \label{tab:low_high}
  \resizebox{1\linewidth}{!}{
  \begin{tabular}{llccc}
    \toprule
    \multirow{2}{*}{Metric} & \multirow{2}{*}{Method} & \multicolumn{3}{c}{Target age} \\
    \cmidrule(lr){3-5}
    & & 1 & 80 & ALL \\
    \midrule
    \multirow{2}{*}{AGE~$\downarrow$} & Ours w/o ``\texttt{baby}"/``\texttt{elderly}" & 13.9 & 10.6 & 12.5 \\
    & Ours & \textbf{9.22} & \textbf{7.07} & \textbf{11.3} \\
    \midrule
    \multirow{2}{*}{ID~$\downarrow$} & Ours w/o ``\texttt{baby}"/``\texttt{elderly}" & \textbf{0.142} & \textbf{0.0944} & \textbf{0.0886} \\
    & Ours & 0.155 & 0.128 & 0.0923 \\
    \bottomrule
  \end{tabular}
  }
\end{table}

\begin{figure}
  \centering
  \includegraphics[width=\linewidth]{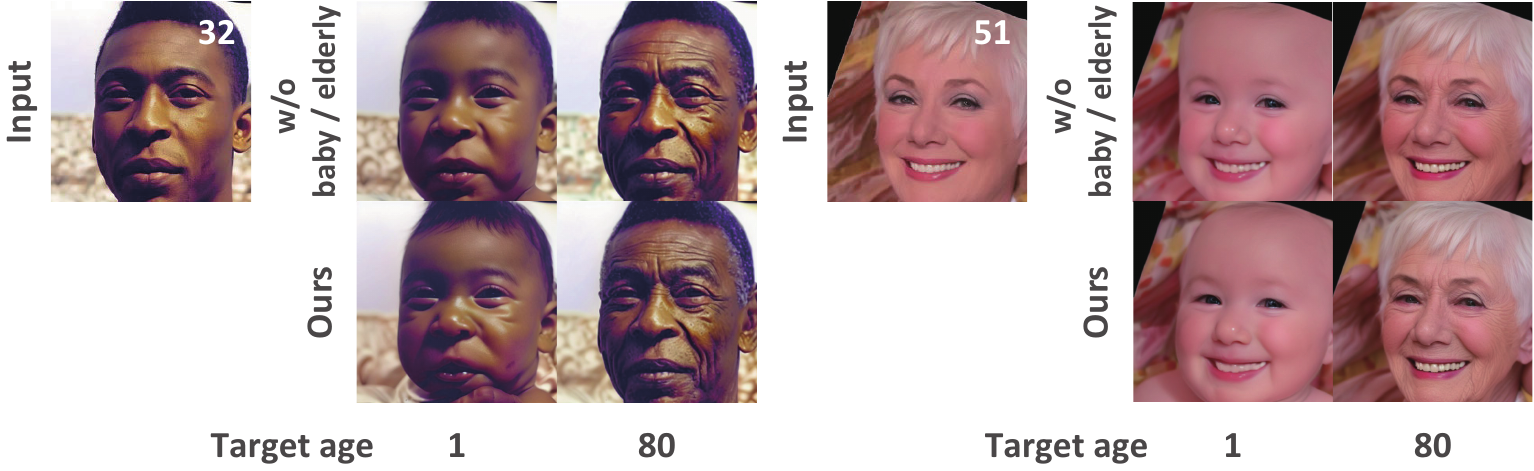}
  \caption{
  Qualitative comparison with and without token replacement for extreme age. 
  }
  \label{fig:low_high}
\end{figure}

Table~\ref{tab:low_high} demonstrates significant improvements in AGE for the target ages (1 and 80) affected by our token replacement, with only a small drop in ID.
Figure~\ref{fig:low_high} shows the qualitative results.
The qualitative analysis also confirms that the method achieves more pronounced age modifications for both lower and higher ages.
For younger ages, facial features become rounder. 
For older ages, a part of the hair changes to white, and wrinkles increase, demonstrating realistic age transformations.

%% file: 10_conclusion.tex
\section{Conclusion}
\label{chap:conclusion}
In this paper, we proposed the first diffusion model-based method for personalized age transformation, which enhances the performance of both age editing and identity preservation. 
Our method fine-tunes a pretrained LDM using self-reference images and their corresponding ages, adapting the model to a specific individual.
Simultaneously, it specializes in age transformations by learning regularization images labeled with fine-grained ages.
To prevent overfitting and ensure stable age transformations, we employed LoRA during training and inference.
Furthermore, we developed effective prompt designs such as modification of age representation, use of self-reference image age, and token replacement for extreme age. 
Quantitative and qualitative evaluations demonstrated that our method achieves age editing performance comparable to state-of-the-art approaches while effectively preserving identity.

\paragraph{Limitations and future work.}
\begin{figure*}[t]
  \centering
  \includegraphics[width=\linewidth]{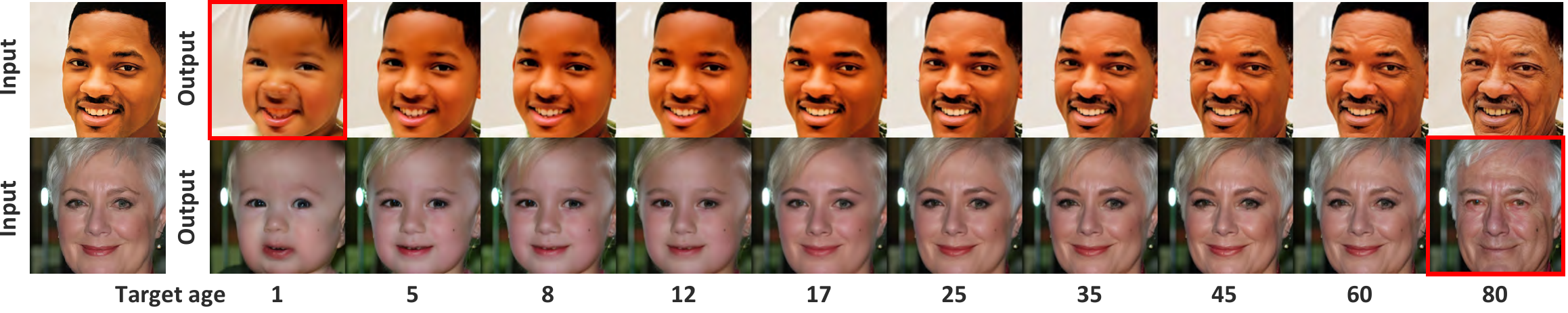}
  \caption{
  Our failure cases (red boxes). In the top row, the shape and color of the teeth, as well as the shadow under the nose, appear unnatural. In the bottom row, the gender changes during age editing when transforming to an older age.
  }
  \label{fig:failed}
\end{figure*}

Figure~\ref{fig:failed} shows failure cases of our method.
One limitation is that artifacts may still occur when performing extreme age transformations, particularly when editing toward younger ages, which require significant shape deformation.
These artifacts are most commonly observed in regions where structural changes are expected, such as the jawline, mouth, and nose.
Additionally, in rare cases, the gender of the subject changes unintentionally during editing. 
This is probably because our method replaces the term ``\texttt{person}'' in the prompt based solely on age without considering gender; we used gender-neutral terms like ``\texttt{baby}" or ``\texttt{elderly}" for extreme age groups, unlike gender-specific terms like ``\texttt{man}" or ``\texttt{woman}". 
Future improvements in prompt design could potentially address this issue and further enhance the reliability of age transformations. 

%% file: 12_appendix.tex
\section*{Appendix}
\label{sec:appendix_section}
%Supplementary material goes here.
\renewcommand{\thesubsection}{\Alph{subsection}}
\setcounter{subsection}{0}
\subsection{Additional Qualitative Comparison}
\begin{figure*}[h]
  \centering
  \includegraphics[height=1.\textheight]{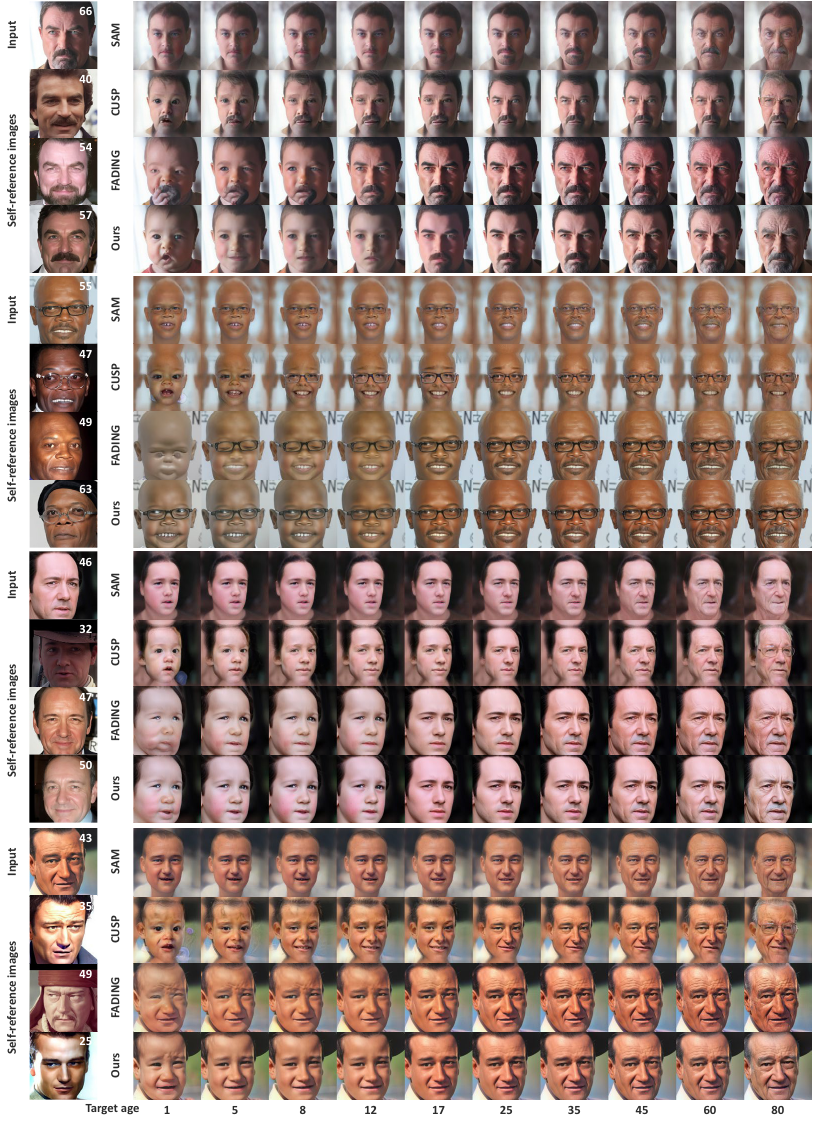}
  \caption{
  Additional qualitative comparison between our method and the existing methods~\cite{alaluf2021only,gomez2022custom,chen2023face}.
  }
  \label{fig:more_1}
\end{figure*}

\begin{figure*}[h]
  \centering
  \includegraphics[height=1.\textheight]{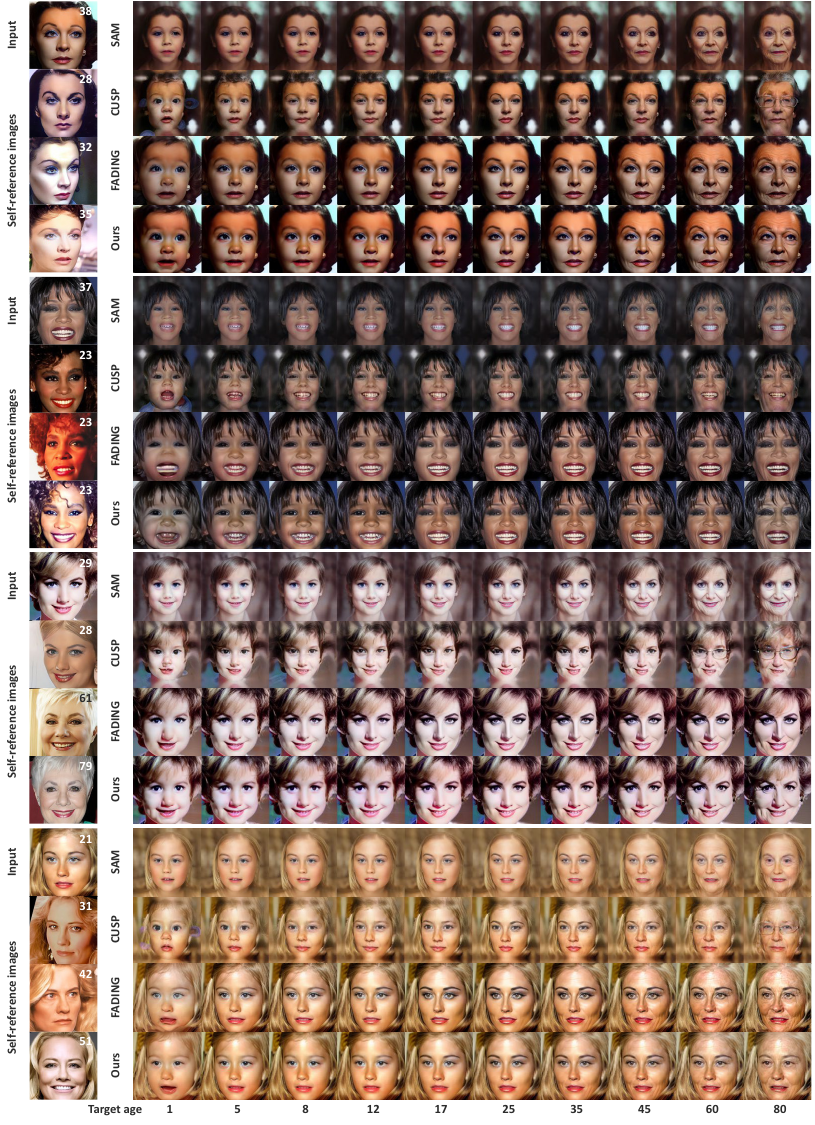}
  \caption{
  Additional qualitative comparison between our method and the existing methods~\cite{alaluf2021only,gomez2022custom,chen2023face}.
  }
  \label{fig:more_2}
\end{figure*}

Figures~\ref{fig:more_1} and \ref{fig:more_2} present additional qualitative comparisons.
SAM~\cite{alaluf2021only} exhibits low inversion accuracy, resulting in inadequate modifications for both younger and older age transformations.
CUSP~\cite{gomez2022custom} demonstrates strong performance in editing for younger ages but struggles to maintain identity preservation across all target ages.
FADING~\cite{chen2023face} maintains identity well due to its high inversion accuracy, but it frequently fails to perform successful age modifications.
In contrast, our method achieves more consistent and stable age editing.
Furthermore, thanks to our prompt design, editing performance for both younger and older ages has significantly improved.
Notably, our method produces natural changes in hair volume and color, which are crucial for realistic age transformations.

